# Electromagnetic force and torque in Lorentz and Einstein-Laub formulations

Masud Mansuripur

College of Optical Sciences, The University of Arizona, Tucson, Arizona 85721

The Lorentz force law of classical electrodynamics requires the introduction of hidden energy and hidden momentum in situations where an electric field acts on a magnetic material. In contrast, the Einstein-Laub formulation does not invoke hidden entities. The total force and torque exerted by electromagnetic fields on a given object are independent of whether the force and torque densities are evaluated using the law of Lorentz or that of Einstein and Laub. Hidden entities aside, the two formulations differ only in their predicted force and torque *distributions* throughout material media.

**1. Introduction**. The classical theory of electrodynamics is built upon Maxwell's equations and the concepts of electromagnetic (EM) field, force, energy, and momentum, which are intimately tied together by Poynting's theorem and the Lorentz force law. Whereas Maxwell's macroscopic equations relate the electric and magnetic fields to their material sources (i.e., charge, current, polarization and magnetization), Poynting's theorem governs the flow of EM energy and its exchange between fields and material media, while the Lorentz law regulates the back-and-forth transfer of momentum between the media and the fields. As it turns out, an alternative force law, first proposed in 1908 by Einstein and Laub, exists that is consistent with Maxwell's macroscopic equations and complies with the conservation laws as well as with the requirements of special relativity. While the Lorentz law requires the introduction of hidden energy and hidden momentum in situations where an electric field acts on a magnetic material, the Einstein-Laub formulation of EM force and torque does not invoke hidden entities under such circumstances. Moreover, the *total* force and the *total* torque exerted by EM fields on any given object turn out to be independent of whether force and torque densities are evaluated using the Lorentz law or obtained in accordance with the Einstein-Laub formulas. Hidden entities aside, the two formulations differ only in their predicted force and torque *distributions* throughout material media. Such differences in distribution are occasionally measurable, and could serve as a guide in deciding which formulation, if either, corresponds to physical reality.

 It has been suggested that magnetic dipoles cannot exist in classical physics. A famous argument in support of this assertion is given in Niels Bohr's PhD dissertation (1911).[1] The argument is now known as the Bohr-van Leeuwen theorem. Feynman also mentions this point briefly in his *Lectures on Physics*.[2] We believe this is a consequence of accepting Maxwell's so-called *microscopic* equations as the fundamental laws of electrodynamics, in which case Ampèrian current loops become the standard model for magnetic dipoles; it is, in fact, the existence and properties of such stable current loops that spell trouble for magnetism in classical physics. If, instead, one adopts Maxwell's *macroscopic* equations as fundamental (i.e., equations that admit, in addition to electric charge and current, both electric and magnetic dipoles as basic building blocks of matter), then classical physics will have incorporated electric and magnetic dipoles without much ado—simply by fiat. One will have to make a substantial adjustment in his/her view of Maxwell's *macroscopic* equations in order to arrive at such a conclusion. Once the adjustment is made, however, classical electrodynamics will be enriched and strengthened. Elsewhere we have shown the self-consistency of this viewpoint as well as its consistency with the conservation laws and with special relativity.[3-8] It is from this perspective that the Einstein-Laub force and torque equations (rather than the Lorentz force law) can be considered as foundational postulates of classical electrodynamics.

In this view, polarization $\boldsymbol{P}(\boldsymbol{r},t)$ and magnetization $\boldsymbol{M}(\boldsymbol{r},t)$ are interpreted as the densities of electric and magnetic dipoles at different locations in space-time, with the implicit assumption that individual dipoles are point-dipoles. That way, one can fill every region of space, no matter how small, with as few or as many dipoles as one may desire. The density of each kind of dipole can thus be an arbitrary function of the position coordinate, $\boldsymbol{r}$.

**2. Maxwell's macroscopic equations**. We begin our examination of Maxwell's macroscopic equations with his first equation, also known as Gauss's law,[9] namely,

$$\boldsymbol{\nabla} \cdot \boldsymbol{D}(\boldsymbol{r},t) = \rho_{\text{free}}(\boldsymbol{r},t). \tag{1}$$

Since, by definition,

$$\boldsymbol{D}(\boldsymbol{r},t) = \varepsilon_0 \boldsymbol{E}(\boldsymbol{r},t) + \boldsymbol{P}(\boldsymbol{r},t), \tag{2}$$

where $\varepsilon_0$ is the permittivity of free space, Eq. (1) may be written as

$$\varepsilon_0 \boldsymbol{\nabla} \cdot \boldsymbol{E} = \rho_{\text{free}} - \boldsymbol{\nabla} \cdot \boldsymbol{P}. \tag{3}$$

Therefore, one is justified in treating $-\boldsymbol{\nabla} \cdot \boldsymbol{P}(\boldsymbol{r},t)$ as the bound-charge-density $\rho_{\text{bound}}(\boldsymbol{r},t)$ associated with electric dipoles. This is also exactly what has traditionally been taken to be the bound charge-density of electric dipoles.[2,9,10] Aside from the fact that we assume $\boldsymbol{P}(\boldsymbol{r},t)$ can represent any arbitrary vector function of space-time, it is clear that the divergence of this function produces a bound charge-density, in the same way that one has always assumed that polarization should. This does *not* mean that the bound charge-density is an arbitrary function of space-time; it is the *divergence* of another (arbitrary) function, but the bound charge-density is not arbitrary. Together with the corresponding bound-current-density, $\boldsymbol{J}_{\text{bound}}(\boldsymbol{r},t) = \partial \boldsymbol{P}/\partial t$ (introduced below), they satisfy the charge-current continuity equation, $\boldsymbol{\nabla} \cdot \boldsymbol{J} + \partial \rho/\partial t = 0$.

Next we examine Maxwell's second (macroscopic) equation, also known as the Maxwell-Ampère equation:[9]

$$\boldsymbol{\nabla} \times \boldsymbol{H} = \boldsymbol{J}_{\text{free}} + \partial \boldsymbol{D}/\partial t. \tag{4}$$

This can be written as

$$\boldsymbol{\nabla} \times \boldsymbol{B} = \mu_0 (\boldsymbol{J}_{\text{free}} + \partial \boldsymbol{P}/\partial t + \mu_0^{-1} \boldsymbol{\nabla} \times \boldsymbol{M}) + (1/c^2) \partial \boldsymbol{E}/\partial t, \tag{5}$$

where, by definition,

$$\boldsymbol{B}(\boldsymbol{r},t) = \mu_0 \boldsymbol{H}(\boldsymbol{r},t) + \boldsymbol{M}(\boldsymbol{r},t). \tag{6}$$

Here $\mu_0$ is the permeability of free space and $c = 1/\sqrt{\mu_0 \varepsilon_0}$ is the speed of light in vacuum. The two contributions to the bound current-density are seen to be $\partial \boldsymbol{P}/\partial t$ and $\mu_0^{-1} \boldsymbol{\nabla} \times \boldsymbol{M}$. Of course the charge-current continuity equation is satisfied because of the way $\partial \boldsymbol{P}/\partial t$ and $-\boldsymbol{\nabla} \cdot \boldsymbol{P}$ are both related to the same (albeit arbitrary) function $\boldsymbol{P}(\boldsymbol{r},t)$, and also because $\boldsymbol{\nabla} \cdot (\boldsymbol{\nabla} \times \boldsymbol{M})$ is always zero. Again, $\boldsymbol{M}(\boldsymbol{r},t)$ may be an arbitrary function of the space-time, but its contribution to bound electrical charge (zero) and bound electrical current ($\mu_0^{-1} \boldsymbol{\nabla} \times \boldsymbol{M}$) satisfy the charge-current continuity equation.

The above arguments reveal that, once one accepts the functions $\boldsymbol{P}(\boldsymbol{r},t)$ and $\boldsymbol{M}(\boldsymbol{r},t)$ appearing in the macroscopic equations as representing arbitrary distributions of electric and magnetic dipole densities, there will be no need to introduce additional bound charges and currents. The vector functions $\boldsymbol{P}(\boldsymbol{r},t)$ and $\boldsymbol{M}(\boldsymbol{r},t)$, precisely because of the way that they appear in Maxwell's macroscopic equations, give rise to the same bound-charge and bound-current densities as have always been associated with polarization and magnetization.

Maxwell's third equation, $\nabla \times \boldsymbol{E} = -\partial \boldsymbol{B}/\partial t$, also known as Faraday's law,[9] does not require any reinterpretation, nor does the fourth equation, $\nabla \cdot \boldsymbol{B} = 0$. Together with Eqs.(3) and (5), these equations relate $\rho_{\text{total}} = \rho_{\text{free}} - \nabla \cdot \boldsymbol{P}$ and $\boldsymbol{J}_{\text{total}} = \boldsymbol{J}_{\text{free}} + \partial \boldsymbol{P}/\partial t + \mu_0^{-1} \nabla \times \boldsymbol{M}$ to $\boldsymbol{E}(\boldsymbol{r},t)$, the electric field at an arbitrary location $(\boldsymbol{r},t)$, and to $\boldsymbol{B}(\boldsymbol{r},t)$, the local, instantaneous value of the magnetic induction, which is related to the $H$-field via Eq.(6). Alternatively, one may write the macroscopic equations in the following, equally valid way:

$$\varepsilon_0 \nabla \cdot \boldsymbol{E} = \rho_{\text{free}} - \nabla \cdot \boldsymbol{P}, \tag{7}$$

$$\nabla \times \boldsymbol{H} = (\boldsymbol{J}_{\text{free}} + \partial \boldsymbol{P}/\partial t) + \varepsilon_0 \partial \boldsymbol{E}/\partial t, \tag{8}$$

$$\nabla \times \boldsymbol{E} = -\partial \boldsymbol{M}/\partial t - \mu_0 \partial \boldsymbol{H}/\partial t, \tag{9}$$

$$\mu_0 \nabla \cdot \boldsymbol{H} = -\nabla \cdot \boldsymbol{M}. \tag{10}$$

Clearly, the above equations now relate the total electric-charge and electric-current densities $\rho_{\text{total}}^{(e)} = \rho_{\text{free}} - \nabla \cdot \boldsymbol{P}$ and $\boldsymbol{J}_{\text{total}}^{(e)} = \boldsymbol{J}_{\text{free}} + \partial \boldsymbol{P}/\partial t$ as well as the total magnetic-charge and magnetic-current densities $\rho_{\text{total}}^{(m)} = -\nabla \cdot \boldsymbol{M}$ and $\boldsymbol{J}_{\text{total}}^{(m)} = \partial \boldsymbol{M}/\partial t$ to $\boldsymbol{E}(\boldsymbol{r},t)$ and $\boldsymbol{H}(\boldsymbol{r},t)$. Neither set of equations is superior to the other. While electric dipoles in both cases are represented by electric charges and currents, magnetic dipoles appear as Ampèrian electric-current-loops in the first, and as Gilbertian magnetic charges and currents in the second case.

**3. Poynting's theorem**. Maxwell's macroscopic equations yield two different, albeit closely related, versions of the Poynting theorem.[2,9-11] The first version is obtained by dot-multiplying $\boldsymbol{E}$ and $\boldsymbol{H}$ into the Maxwell-Ampère and Faraday equations, respectively, then subtracting one equation from the other. We will have

$$\nabla \cdot (\boldsymbol{E} \times \boldsymbol{H}) + \tfrac{\partial}{\partial t}(\tfrac{1}{2}\varepsilon_0 \boldsymbol{E} \cdot \boldsymbol{E} + \tfrac{1}{2}\mu_0 \boldsymbol{H} \cdot \boldsymbol{H}) + \boldsymbol{E} \cdot \left(\boldsymbol{J}_{\text{free}} + \tfrac{\partial \boldsymbol{P}}{\partial t}\right) + \boldsymbol{H} \cdot \tfrac{\partial \boldsymbol{M}}{\partial t} = 0. \tag{11}$$

According to the above version of the theorem, the Poynting vector is $\boldsymbol{S}_1 = \boldsymbol{E} \times \boldsymbol{H}$, the stored energy-density in the fields is $\mathcal{E}_1 = \tfrac{1}{2}\varepsilon_0 \boldsymbol{E} \cdot \boldsymbol{E} + \tfrac{1}{2}\mu_0 \boldsymbol{H} \cdot \boldsymbol{H}$, and the rates of EM energy exchange between fields, on the one hand, and free currents and electric and magnetic dipoles, on the other hand, are $\boldsymbol{E} \cdot \boldsymbol{J}_{\text{free}}$, $\boldsymbol{E} \cdot \partial \boldsymbol{P}/\partial t$, and $\boldsymbol{H} \cdot \partial \boldsymbol{M}/\partial t$, respectively.

To obtain the second version of Poynting's theorem, we dot-multiply $\boldsymbol{E}$ and $\boldsymbol{B}$ into the Maxwell-Ampère and Faraday equations, respectively, then subtract one from the other to find[11]

$$\nabla \cdot (\mu_0^{-1} \boldsymbol{E} \times \boldsymbol{B}) + \tfrac{\partial}{\partial t}(\tfrac{1}{2}\varepsilon_0 \boldsymbol{E} \cdot \boldsymbol{E} + \tfrac{1}{2}\mu_0^{-1} \boldsymbol{B} \cdot \boldsymbol{B}) + \boldsymbol{E} \cdot \left(\boldsymbol{J}_{\text{free}} + \tfrac{\partial \boldsymbol{P}}{\partial t} + \mu_0^{-1} \nabla \times \boldsymbol{M}\right) = 0. \tag{12}$$

Here the Poynting vector is $\boldsymbol{S}_2 = \mu_0^{-1} \boldsymbol{E} \times \boldsymbol{B}$, the stored energy-density in the fields is $\mathcal{E}_2 = \tfrac{1}{2}\varepsilon_0 \boldsymbol{E} \cdot \boldsymbol{E} + \tfrac{1}{2}\mu_0^{-1} \boldsymbol{B} \cdot \boldsymbol{B}$, and the rates of EM energy exchange between the fields and the material media are $\boldsymbol{E} \cdot \boldsymbol{J}_{\text{free}}$, $\boldsymbol{E} \cdot \partial \boldsymbol{P}/\partial t$, and $\boldsymbol{E} \cdot \mu_0^{-1} \nabla \times \boldsymbol{M}$.

There are obvious differences between the above versions of Poynting's theorem. The magnetic field energy is stored in the $H$-field in the first version, and in the $B$-field in the second. In both versions, the $E$-field acts on free currents and on electric dipoles to exchange energy with material media. However, whereas in the first version the $H$-field acts on the *magnetic* current $\partial \boldsymbol{M}/\partial t$ to exchange energy with magnetic dipoles, it is the $E$-field in the second version that acts on the corresponding *electric* current-density $\mu_0^{-1} \nabla \times \boldsymbol{M}$.

Another significant difference between the two versions of the theorem is in the Poynting vector itself, where $S_1 - S_2 = \mu_0^{-1} M \times E$. This difference is associated with the rate of flow (per unit area per unit time) of the so-called *hidden energy*. To appreciate the role of hidden energy in EM systems, consider a plane-wave at normal incidence from vacuum onto the flat surface of a magnetic dielectric specified by its relative permittivity $\varepsilon$ and relative permeability $\mu$; see Fig. 1. Maxwell's equations impose continuity constraints on the tangential $E$ and $H$ components at the boundary. Consequently, while $S_1$ remains continuous at the boundary, there will be a discontinuity in $S_2$, simply because $B_\parallel$ at the interface is discontinuous. The discontinuity in the energy flux at the boundary (when $S_2$ is taken to be the Poynting vector) can be accommodated by taking into account the hidden energy flux $\mu_0^{-1} M \times E$.

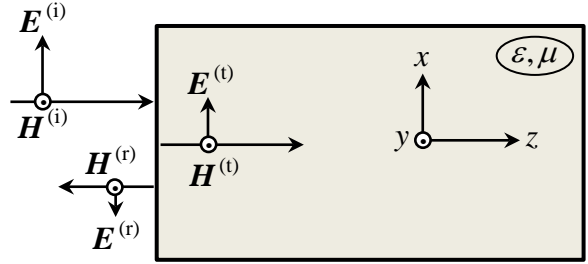

**Fig. 1**. A magnetic dielectric slab, having permittivity $\varepsilon$ and permeability $\mu$, is illuminated by a plane electromagnetic wave at normal incidence. The continuity of the tangential components of the $E$ and $H$ fields at the entrance facet, namely, $E_x^{(t)} = E_x^{(i)} + E_x^{(r)}$ and $H_y^{(t)} = H_y^{(i)} + H_y^{(r)}$, guarantees the continuity of the energy flux vector $S_1 = E \times H$. By the same token, the discontinuity of the tangential component of the $B$ field, namely, $B_y^{(t)} \neq B_y^{(i)} + B_y^{(r)}$, indicates that $S_2 = \mu_0^{-1} E \times B$ is discontinuous at the entrance facet.

**4. Electromagnetic force and torque according to Lorentz**. The force law of Lorentz, $f = q(E + V \times B)$, is the expression of the force $f$ exerted on a point-charge $q$ moving with velocity $V$ in the external electric and magnetic fields $E(r,t)$ and $B(r,t)$. A straightforward generalization yields[2,9-11]

$$F_L(r,t) = \rho(r,t)E(r,t) + J(r,t) \times B(r,t), \quad (13a)$$

$$T_L(r,t) = r \times F_L(r,t). \quad (13b)$$

where $F_L$ is the Lorentz force-density, $T_L$ is the corresponding torque-density, $\rho$ is electric-charge-density, and $J$ is electric-current-density. There is no *a priori* reason to distinguish between free and bound charges, nor between free and bound currents. In other words, $\rho(r,t)$ and $J(r,t)$ in the above Lorentz formulation could arise from various combinations of free and bound electric charges. Thus, in accordance with Eqs. (3) and (5), the total electric charge-density $\rho_{total} = \rho_{free} - \nabla \cdot P$, and the total electric current-density $J_{total} = J_{free} + \partial P/\partial t + \mu_0^{-1} \nabla \times M$, give rise to a net force-density $F_L = \rho_{total} E + J_{total} \times B$.

In conjunction with Maxwell's macroscopic equations, the Lorentz force-density of Eq. (13a) leads to the Maxwell-Lorentz stress tensor and the Livens EM momentum-density[9,12,13]

$$\overleftrightarrow{\mathcal{T}}_L(r,t) = \tfrac{1}{2}(\varepsilon_0 E \cdot E + \mu_0^{-1} B \cdot B)\overleftrightarrow{I} - \varepsilon_0 EE - \mu_0^{-1} BB, \quad (14a)$$

$$\wp_L(r,t) = \varepsilon_0 E \times B, \quad (14b)$$

which satisfy the following continuity equation:

$$\overleftrightarrow{\nabla} \cdot \overleftrightarrow{\mathcal{T}}_L + \frac{\partial \pmb{p}_L}{\partial t} + \pmb{F}_L = 0. \tag{14c}$$

Equation (14c) is the statement of conservation of linear momentum, just as Eq.(12) is the corresponding statement of conservation of energy. According to Eq.(14c), the force-density $\pmb{F}_L(\pmb{r},t)$ is the rate of exchange of momentum between EM fields and material media. Note that the Livens momentum-density of Eq.(14b) is related to the Poynting vector appearing in Eq.(12) via $\pmb{p}_L(\pmb{r},t) = \pmb{S}_2(\pmb{r},t)/c^2$.

A well-known thought experiment due to Balazs[14] indicates that the EM field should have the Abraham momentum-density $\pmb{p}_A = \pmb{S}_1/c^2 = \pmb{E} \times \pmb{H}/c^2$. This means that $\pmb{p}_L$ in Eq.(14c) should be replaced by $\pmb{p}_A$, and the Lorentz force-density should be augmented by the difference between the Livens and Abraham momenta, that is, the *actual* force-density should be given by $\pmb{F}_L - \partial(\varepsilon_0 \pmb{M} \times \pmb{E})/\partial t$. The *hidden* momentum-density trapped inside magnetic materials is thus defined as $\pmb{p}_{\text{hidden}}(\pmb{r},t) = \varepsilon_0 \pmb{M} \times \pmb{E}$. In other words, whenever magnetic dipoles are subjected to an $E$-field, the time-rate-of-change of hidden momentum must be subtracted from the Lorentz force-density $\pmb{F}_L$ in order to arrive at the *actual* force exerted on the material medium.[15-45]

**5. Alternative perspective on classical electrodynamics**. In preparation for a discussion of the Einstein-Laub formalism, we need to explain a few facts about Maxwell's macroscopic equations. Some readers may find the material covered in the present section trivial. However, many physicists appear to be deeply entrenched in the standard Maxwell-Lorentz theory, to the extent that they find the Einstein-Laub approach disturbing. Our preliminary suggestions in support of the Einstein-Laub formalism may be summarized as follows: Accept Maxwell's macroscopic equations *exactly* as they are written, that is, as precise relations among several scalar and vector functions of space-time, with the implicit assumption that $\pmb{P}(\pmb{r},t)$ and $\pmb{M}(\pmb{r},t)$ may or may not be subjugated to $\pmb{E}(\pmb{r},t)$ and $\pmb{H}(\pmb{r},t)$ via certain constitutive relations. Do not change the equations, but do not bring along any historical baggage in the form of traditional interpretations. Thus $\rho_{\text{free}}$ and $\pmb{J}_{\text{free}}$ are the densities of free charge and free current, exactly as they have always been understood. $\pmb{P}(\pmb{r},t)$, which appears in the expression for $\pmb{D}(\pmb{r},t)$ in Eq.(2), should be thought of as the density of electric dipoles at the point $(\pmb{r},t)$ in space-time. The traditionalist would take a small volume of space containing hundreds or thousands of atomic dipoles; average the dipole moments contained within this volume, then call that average the local dipole-moment-density $\pmb{P}(\pmb{r},t)$. We suggest to do away with local averaging and its associated historical baggage. Simply stated, we propose to accept the vector function $\pmb{P}(\pmb{r},t)$ as the density of electric point-dipoles at $(\pmb{r},t)$. Thus, even a single point-dipole $\pmb{p}(t)$, located at $\pmb{r}_0 = (x_0, y_0, z_0)$, would have its well-defined polarization density, as follows:

$$\pmb{P}(\pmb{r},t) = \pmb{p}(t)\delta(\pmb{r}-\pmb{r}_0) = \pmb{p}(t)\delta(x-x_0)\delta(y-y_0)\delta(z-z_0). \tag{15}$$

A similar concession is necessary with regard to magnetization $\pmb{M}(\pmb{r},t)$: When one writes $\pmb{B} = \mu_0 \pmb{H} + \pmb{M}$, the historical baggage that comes with the conventional definition of $\pmb{M}(\pmb{r},t)$ must be dispensed with, allowing $\pmb{M}(\pmb{r},t)$ to simply represent the local density of magnetic dipole moments in space-time—no local averaging is needed, and no interpretation in terms of Ampèrian current loops or Gilbertian pairs of magnetic monopoles need be brought along.[45] After all, the macroscopic equations of Maxwell have been extremely successful in describing a range of phenomena involving polarized (or polarizable) and magnetized (or magnetizable) media. Why shouldn't we dispense with the conceptual baggage that is the notion of locally-averaged dipole moment, and allow $\pmb{P}(\pmb{r},t)$ and $\pmb{M}(\pmb{r},t)$ to stand for *idealized* densities of dipole moments at each point in space-time? The equations do not change; their solutions do not change; the free and bound charge and current densities remain precisely what they have been all along. All we are proposing here is to free $P$ and $M$ from their historical straightjacket and allow them to stand on their own. This, by the way, is no different than what has traditionally been done with $\rho_{\text{free}}$ and $J_{\text{free}}$, which are taken to be continuous, differentiable functions of $(r, t)$, even though, in physical reality, charge and current are granular and fluctuating at some microscopic level.

We have already discussed, in sections 2 and 3, the different interpretations of $P$ and $M$ within Maxwell's macroscopic equations. Here we would like to emphasize once again that such interpretations do *not* change the equations at all: Same equations, same solutions, same sources of the EM fields, and same fields. The only things that are different now are: (i) the notion of local averaging attached to the dipole densities is being abandoned, and (ii) no specific models such as pair-of-charges-on-a-stick or Ampèrian current-loops are being assumed for the dipoles.

In Section 6 we will describe EM force and torque according to Einstein and Laub. The differences between the Lorentz and Einstein-Laub formalisms are significant in some respects but not so great in others. We must focus on the equations and what they actually say—rather than what people may have read into these equations in the past. There is a historical parallel here: When Maxwell formulated his theory, he had in mind certain notions of what the ether looked like; there were all sorts of wheels and gears that turned this way and that way in order to sustain the EM fields. Later, people decided that what mattered most were the equations themselves. Heinrich Hertz famously said that Maxwell's theory is Maxwell's equations. Feynman said that, when the scaffolding is removed, the magnificent edifice that Maxwell built stands on its own. What we are promoting here is a similar view of the *macroscopic* equations of Maxwell. Let us take the equations at face value and allow the various vector and scalar functions appearing in these equations to represent (idealized) elements of reality, with no Ampèrian or Gilbertian or Lorentzian models and interpretations lurking in the background.

Of course, everything happens in vacuum. There are four sources of the EM field residing in vacuum: electric charge, electric current, electric dipoles, and magnetic dipoles. The densities of these are given by $\rho_{\text{free}}(r,t)$, $J_{\text{free}}(r,t)$, $P(r,t)$ and $M(r,t)$. The first is a scalar function, the other three are vector functions. The charge and current densities satisfy the continuity equation. $P(r,t)$ and $M(r,t)$ are arbitrary, but when written as bound electric charge and current densities, namely, $-\nabla \cdot P$ and $\partial P/\partial t + \mu_0^{-1} \nabla \times M$, they also automatically satisfy the continuity equation. Material media are nothing but collections of one or more of these four sources.

Adopting the new perspective on classical electrodynamics requires that one remain vigilant and mindful of the aforementioned historical baggage, which appears in different guises and has several components, as follows:

i) $P$ and $M$ are traditionally taken to be local averages of dipole densities; this is not necessary. For example, a single electron in free space, moving at some velocity $V$, is a point-charge as well as a localized current-density. The electron also carries a magnetic dipole moment (due to its spin), and is accompanied by a relativistically-induced electric dipole moment. All four sources are thus present at one point of space at any given time. The concept of local averaging obviously does not apply to this single point-particle. Nevertheless, Maxwell's macroscopic equations (i.e., those relating $E, D, B, H, \rho_{\text{free}}, J_{\text{free}}$) apply to this situation, as well as to any other situation involving arbitrary distributions of the four EM sources; no local averaging is required.

ii) The $E$ and $B$ fields are assumed to be fundamental, while $H$ and $D$ are said to be *derived* or *secondary* fields; again, there is no need to make such distinctions. Of course there is a strict relationship among these four fields, namely, $D = \varepsilon_0 E + P$ and $B = \mu_0 H + M$. So, knowledge of $E$ and $B$ together with that of $P$ and $M$ is sufficient to uniquely determine $D$ and $H$. Also, the ($E, B$) pair forms a 2$^{nd}$ rank tensor that can be Lorentz-transformed between different inertial frames. However, the ($D, H$) pair also forms a 2$^{nd}$ rank tensor which is similarly transformed.[46] There is no logical need to restrict our worldview to only two sources (electric charge and electric current) and two fields ($E$ and $B$). We can keep an open mind and accept that Maxwell's equations are exact (mathematical) relations among four sources ($\rho_{\text{free}}, J_{\text{free}}, P, M$) and four fields ($E, D, H, B$). This may not appeal to our minimalist sense of beauty and elegance; at first sight, it also appears to violate the dictum of Occam's razor, but there is no logical inconsistency in accepting this larger set of fields and sources as containing the building blocks of classical electrodynamics.

Let us emphasize that eliminating the notion of *local averaging* from the macroscopic equations is an important step toward accepting this enlarged system of fields and sources as *fundamental*, and the corresponding Maxwell equations as *exact* relations among the fields and their sources. The notion of local averaging gives the impression that there is something approximate or inaccurate about Maxwell's *macroscopic* equations, that they are perhaps not valid on small (atomic) scales. However, mathematically speaking, there is nothing approximate about the macroscopic equations. If we allow $\rho_{\text{free}}, J_{\text{free}}, P, M, E, D, B, H$ to be continuous and differentiable functions of space-time, then the *macroscopic* equations relate these functions to each other at all scales (spatial as well as temporal). Of course, the real (physical) world is discrete, and the spatial dimensions may not extend below a certain scale, and Heisenberg's uncertainty principle puts a limit on the accuracy with which our equations can describe the real world, and there are thermal and quantum fluctuations. Nevertheless, we are interested here in classical (as opposed to quantum) electrodynamics, and in Maxwell's equations as *mathematical* relations describing an ideal (continuous, differentiable) world.

iii) $P$ and $M$ are assumed to be reducible to bound electric charge and current densities; something that is possible, but not necessary. Of course, one can *always* take the macroscopic equations and eliminate $D$ and $H$ from them. One will then end up with a total electric charge-density, $\rho_{\text{free}} - \nabla \cdot P$, and a total electric current density, $J_{\text{free}} + \partial P / \partial t + \mu_0^{-1} \nabla \times M$. This may inform us that the EM world consists solely of electrical charges and currents, and that $E$ and $B$ are the only relevant fields. One can then bring in the Lorentz force law, apply it to the *total* charge and *total* current densities thus obtained, and claim that everything is self-consistent, manifestly covariant, consistent with the conservation laws, etc. Needless to say, one now needs to introduce the notions of hidden energy and hidden momentum in conjunction with magnetic dipoles, because $\mu_0^{-1} \nabla \times M$ requires special treatment in the above formalism. However, as Griffiths and Hnizdo clearly show,[45] this is a necessary consequence of treating magnetic dipoles as Ampèrian current loops—that is, *assuming* that $M(r,t)$ is associated with a bound electric current-density, $\mu_0^{-1} \nabla \times M$.

One advantage of accepting the macroscopic equations as *fundamental* is that we will no longer be locked into accepting the aforementioned formalism. For instance, by deleting $D$ and $B$ from the equations, we find that magnetization can be *equivalently* expressed as a *magnetic* charge-density, $-\nabla \cdot M$, and a magnetic current-density, $\partial M / \partial t$; see Sec. 2. This is apparently what some authors have in mind when they refer to the Gilbert model, but it is really nothing different than the Ampèrian model. (We are *not* talking here about magnetic monopoles; just whatever it is that $-\nabla \cdot \boldsymbol{M}$ and $\partial \boldsymbol{M}/\partial t$ represent within Maxwell's equations.) Rewriting the macroscopic equations in different ways does *not* change anything. We still have the same equations, same solutions, same sources, same fields, etc. Only now we are *interpreting* the sources differently.

Even the so-called "contact term" for a Gilbert dipole is the same as that for an Ampèrian-loop dipole. This is because inside a uniformly-magnetized spherical particle of magnetization $\boldsymbol{M}_0$, the fields are $\boldsymbol{H}_0 = -\boldsymbol{M}_0/(3\mu_0)$ and $\boldsymbol{B}_0 = \tfrac{2}{3}\boldsymbol{M}_0$.[9] To get to a point-dipole, assume that the spherical particle shrinks in size, in which case the contact term becomes $\tfrac{2}{3}\boldsymbol{M}_0 \delta(\boldsymbol{r})$ for the Ampèrian-loop dipole and $-\tfrac{1}{3}(\boldsymbol{M}_0/\mu_0)\delta(\boldsymbol{r})$ for the Gilbertian dipole. Appearances to the contrary notwithstanding, these contact terms are identical, considering that one represents the internal $B$-field while the other represents the internal $H$-field. Consequently, both the external and internal fields of the dipole turn out to be the same, whether one represents the magnetic dipoles with $\mu_0^{-1}\nabla \times \boldsymbol{M}$ (Ampèrian) or with $-\nabla \cdot \boldsymbol{M}$ and $\partial \boldsymbol{M}/\partial t$ (Gilbertian). Indeed, there *cannot* be any differences between the two models, because in both cases we are solving the same (macroscopic) equations; re-arranging the various terms within a system of equations cannot change the solutions.

The Lorentz law applies to total (i.e., free plus bound) electrical charge and electrical current densities. In the presence of magnetic media, the Lorentz formalism requires the notions of hidden energy and hidden momentum in order to remain compatible with special relativity and with the conservation laws. In contrast, the Einstein-Laub force and torque equations satisfy the requirements of special relativity and the conservation laws *without* the need for hidden entities.[3,6,7,8,13,47,48] (There exist other differences between the Lorentz and Einstein-Laub laws, which we have discussed elsewhere[49] and do not need to repeat here.) The point to be emphasized here is that, by accepting the macroscopic equations of Maxwell (which incorporate four sources and four fields), we allow the consideration of force and torque laws other than those of Lorentz. In doing so, we adhere to 150 years of classical physics, but allow for a broad examination of the concepts of force, torque, energy, momentum, and angular momentum in classical electrodynamics.

**6. Classical electrodynamics according to Einstein and Laub**. In a nutshell, the theory of Einstein and Laub may be described as follows: Maxwell's macroscopic equations + Version 1 of Poynting's theorem + Einstein-Laub laws of force-density and torque-density,[50] namely,

$$\boldsymbol{F}_{EL}(\boldsymbol{r},t) = \rho_{\text{free}}\boldsymbol{E} + \boldsymbol{J}_{\text{free}} \times \mu_0 \boldsymbol{H} + (\boldsymbol{P} \cdot \nabla)\boldsymbol{E} + \frac{\partial \boldsymbol{P}}{\partial t} \times \mu_0 \boldsymbol{H} + (\boldsymbol{M} \cdot \nabla)\boldsymbol{H} - \frac{\partial \boldsymbol{M}}{\partial t} \times \varepsilon_0 \boldsymbol{E}, \quad (16\text{a})$$

$$\boldsymbol{T}_{EL}(\boldsymbol{r},t) = \boldsymbol{r} \times \boldsymbol{F}_{EL}(\boldsymbol{r},t) + \boldsymbol{P} \times \boldsymbol{E} + \boldsymbol{M} \times \boldsymbol{H}. \quad (16\text{b})$$

Maxwell's equations tell how the sources $(\rho_{\text{free}}, \boldsymbol{J}_{\text{free}}, \boldsymbol{P}, \boldsymbol{M})$ produce the fields $(\boldsymbol{E}, \boldsymbol{D}, \boldsymbol{H}, \boldsymbol{B})$. The Poynting theorem tells how energy is exchanged between fields and material media (which consist of the aforementioned sources). The Einstein-Laub formulas describe the mechanical force and torque exerted by fields on material media (which, once again, consist of the aforementioned sources).

Many physicists tend to treat the $\boldsymbol{H}$ field as a second-class citizen, believing that it has been invented solely for the convenience of engineers; the real, fundamental magnetic field, they maintain, is $\boldsymbol{B}$. In contrast, Einstein treats the $(\boldsymbol{H}, \boldsymbol{D})$ pair with the same level of respect and affection that he accords the $(\boldsymbol{E}, \boldsymbol{B})$ pair.[46,50]

In discussing the fundamentals, one need not consider the constitutive relations; they are what they are.[8] Whether they are determined experimentally or theoretically, the Einstein-Laub theory does not question the validity of the constitutive relations; they are simply used in Maxwell's equations to solve for the fields. Also, the Einstein-Laub theory does not modify the relativistic treatment of EM fields and sources. The $(\rho_{\text{free}}, \boldsymbol{J}_{\text{free}})$ pair remains a 4-vector, the $(\boldsymbol{P}, \boldsymbol{M})$ pair, the $(\boldsymbol{E}, \boldsymbol{B})$ pair, and the $(\boldsymbol{D}, \boldsymbol{H})$ pair transform between different inertial frames as $2^{\text{nd}}$ rank tensors, as usual.

That is the entire theory! That is how Einstein and Laub conceive "primordial" electric and magnetic dipoles; there is nothing else to their theory. The standard electrodynamics theory thus differs from that of Einstein and Laub in two respects: (i) the laws of force and torque are those of Lorentz applied to total (i.e., free + bound) electric-charge and electric-current densities, and (ii) Poynting's theorem is derived from Maxwell's *microscopic* equations, resulting in version 2 of the theorem, as discussed in Sec.3.

In conjunction with Maxwell's macroscopic equations, the force-density of Eq.(16a) leads to the Einstein-Laub stress tensor and the Abraham EM momentum-density[12,13]

$$\overleftrightarrow{\boldsymbol{\mathcal{T}}}_{EL}(\boldsymbol{r},t) = \tfrac{1}{2}(\varepsilon_0 \boldsymbol{E}\cdot\boldsymbol{E} + \mu_0 \boldsymbol{H}\cdot\boldsymbol{H})\overleftrightarrow{\boldsymbol{I}} - \boldsymbol{D}\boldsymbol{E} - \boldsymbol{B}\boldsymbol{H}, \tag{17a}$$

$$\boldsymbol{\mathcal{p}}_A(\boldsymbol{r},t) = \boldsymbol{E}\times\boldsymbol{H}/c^2, \tag{17b}$$

which satisfy the following continuity equation:

$$\overleftrightarrow{\boldsymbol{\nabla}}\cdot\overleftrightarrow{\boldsymbol{\mathcal{T}}}_{EL} + \frac{\partial \boldsymbol{\mathcal{p}}_A}{\partial t} + \boldsymbol{F}_{EL} = 0. \tag{17c}$$

Equation (17c) is the statement of conservation of linear momentum, just as Eq.(11) is the corresponding statement of conservation of energy. According to Eq.(17c), the force-density $\boldsymbol{F}_{EL}(\boldsymbol{r},t)$ is the rate of exchange of momentum between EM fields and material media. Note that the Abraham momentum-density of Eq.(17b) is related to the Poynting vector appearing in Eq.(11) via $\boldsymbol{\mathcal{p}}_A(\boldsymbol{r},t) = \boldsymbol{S}_1(\boldsymbol{r},t)/c^2$.

In free-space, where $\boldsymbol{D} = \varepsilon_0 \boldsymbol{E}$ and $\boldsymbol{B} = \mu_0 \boldsymbol{H}$, the stress tensors of Lorentz and Einstein-Laub are identical, that is, $\overleftrightarrow{\boldsymbol{\mathcal{T}}}_L = \overleftrightarrow{\boldsymbol{\mathcal{T}}}_{EL}$. This means that the total EM force exerted on an object surrounded by vacuum will turn out to be the same, irrespective of whether it is computed by integrating the Einstein-Laub force-density $\boldsymbol{F}_{EL}(\boldsymbol{r},t)$ or the *reduced* Lorentz force-density, $\boldsymbol{F}_L(\boldsymbol{r},t) - \partial(\varepsilon_0 \boldsymbol{M}\times\boldsymbol{E})/\partial t$, over the volume of the object.

The absence of hidden momentum from the Einstein-Laub theory is an obvious advantage over the theory of Lorentz. This, however, may be considered a matter of convenience rather than necessity. The real advantage of the Einstein-Laub theory is in the subtle features of force and torque *distributions* inside material media; features that should be accessible to experimental verification in deformable bodies.[49] To appreciate this aspect of the theory consider Fig.2, which compares the action of an external $E$-field on an electric dipole in accordance with the Einstein-Laub and Lorentz interpretations. Atoms and molecules define the natural scale at which real-world dipoles behave as distinct entities—as opposed to ideal dipoles consisting of a pair of point-charges on a stick. As far as Maxwell's equations are concerned, the Einstein-Laub dipoles are indistinguishable from the Lorentz dipoles; they can be represented by ordinary electrical charges and currents, which obey all four of Maxwell's equations. The only difference between the Lorentz and Einstein-Laub dipoles is the way in which they participate in energy, force, and torque laws. While the former dipoles reduce to bound electric charges and currents and obey the Lorentz law, the latter appear as $\boldsymbol{P}, \boldsymbol{M}$, and time-derivatives of $\boldsymbol{P}$ and $\boldsymbol{M}$ in the Einstein-Laub equations for force and torque. As mentioned earlier, the *total* force and *total* torque exerted on a given object are precisely the same in the Lorentz and Einstein-Laub formulations; however, the *distributions* of force and torque predicted by the two formulations can be substantially different.

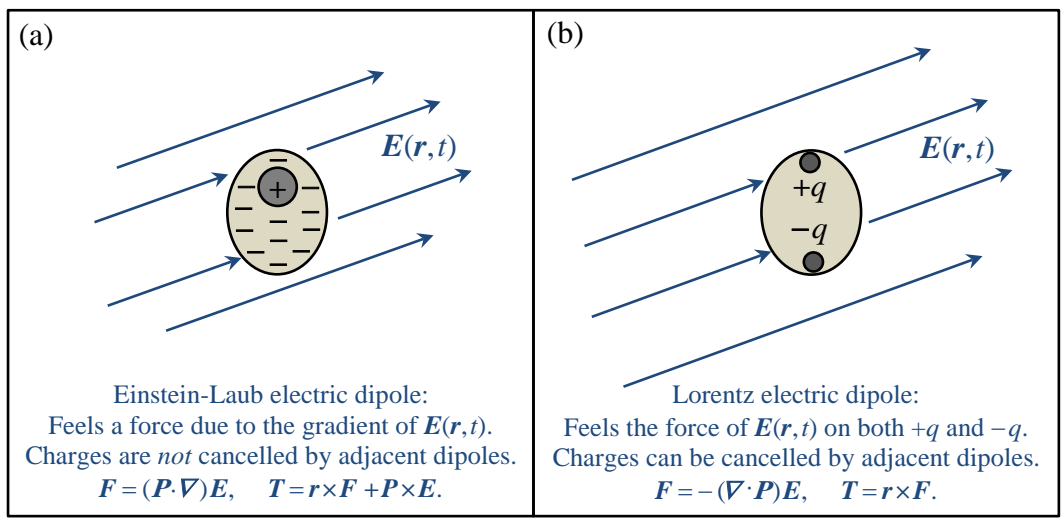

**Fig. 2.** The bounding ellipse represents an individual atom or molecule. In (a) all the minus signs within the ellipse represent a single electron, or perhaps a few electrons, indicating that the electrons of an atom are not localized point-particles; rather they form a cloud surrounding the nucleus. A real-world atomic dipole inside a material medium resembles that depicted in (a). This kind of dipole is "autonomous," in the sense that its charges are not cancelled out by those of the neighboring dipoles; what happens inside the dipole remains inside the dipole. That is why the Einstein-Laub force-density $(\boldsymbol{P}\cdot\boldsymbol{\nabla})\boldsymbol{E}$ is appropriate for describing its interaction with the *E*-field. In contrast, the dipole depicted in (b) is the commonly-imagined, idealized atomic dipole: a pair of point-charges on a stick (or connected via a spring). The charges of such dipoles are cancelled out by the neighboring dipoles when, for example, all the dipoles are lined up vertically. This type of dipole is best described by the Lorentz force-density expression $-(\boldsymbol{\nabla}\cdot\boldsymbol{P})\boldsymbol{E}$. Interestingly, Maxwell's macroscopic equations do *not* distinguish between these two types of dipole; the equations produce unique $\boldsymbol{E}$ and $\boldsymbol{B}$ fields in response to a given function $\boldsymbol{P}(\boldsymbol{r},t)$. The differences between the two types of dipole emerge only in the force and torque equations.

One may wonder whether quadrupoles, octupoles, etc., should also be treated differently. Multi-poles, which apparently are important in nuclear physics, have not been of much use in radiation-pressure studies. Presumably they can be incorporated into Maxwell's macroscopic equations, just as electric and magnetic dipoles have been introduced into these equations. As far as $\boldsymbol{E}$ and $\boldsymbol{B}$ fields are concerned, multi-poles should behave similarly to their corresponding electric charges and currents. However, the force and torque densities associated with the action of fields on primordial multi-poles could differ from those predicted by a straightforward application of the Lorentz law. One will have to examine the experimental evidence of radiation pressure on multi-polar matter in order to arrive at the correct force and torque density equations.

**7. Nature of electric and magnetic dipoles**. In classical physics, atoms and molecules cannot exist and electrons do not have spin. Once the rules of quantum physics succeed in producing atoms, molecules, and spinning electrons, it is only natural to consider electric and magnetic dipoles as primordial (or sovereign) entities which can be imported into classical electro-dynamics—just as other sovereign entities (charge and current) have been brought into the microscopic Maxwell equations. No one has ever seen the inside of an electron, so it is a bit presumptuous to assume that the magnetic dipoles are (in every respect) equivalent to current loops. We should feel lucky that real-world dipoles behave as charges and currents when it comes to producing the $E$ and $B$ fields in accordance with Maxwell's equations—although, as pointed out earlier, the dipoles could as well consist of *magnetic* charges and *magnetic* currents, and continue to produce precisely the same $E$ and $B$ fields via the same Maxwell's equations. Perhaps we push our luck a bit too far when we presume that these primordial dipoles continue to behave as charges and currents in accordance with the Lorentz law—that is, when in our idealized dipole models, we treat electric dipoles as pairs of charges connected via a spring, and magnetic dipoles as small, indestructible current loops.

It may be easier to think of magnetic, rather than electric, dipoles as primordial, because the spin magnetic moment could be an intrinsic property of an unstructured point-particle. In contrast, one would always look at the atom and see, in his/her mind's eye, the positive and negative charges in different locations. This may lead one to consider the pair-of-charges-on-a-spring as a realistic model for the electric dipole. However, consider a single electron traveling with some velocity $V$ in free space. If the magnetic moment $m$ (due to spin) happens to have a component perpendicular to the direction of motion, a stationary observer will see an electric dipole $p = \varepsilon_0 V \times m$ accompanying the magnetic dipole. This electric dipole has no internal structure. If one were inclined to think of electric dipoles as primordial, perhaps this example could provide the license. Be it as it may, the quantum-mechanical nature of atoms and molecules provides equally solid grounds for treating atomic/molecular electric dipoles as well as (orbital- and spin-related) magnetic dipoles as primordial objects.

Magnetism, in particular, creates difficulties for the classical theory. Niels Bohr's Ph.D. thesis has a section in which he proves that diamagnetism and paramagnetism cannot exist in classical physics;[1] this is now known as the Bohr-van Leeuwen theorem. Also, in *The Feynman Lectures on Physics* there is a section discussing why magnetism is forbidden in the classical theory.[2] We believe that accepting Maxwell's *macroscopic* equations as fundamental, reading the Poynting theorem in light of the macroscopic equations (i.e., version 1 of the theorem discussed in Sec. 3), and recognizing the Einstein-Laub formulas as the fundamental laws of force and torque, will open the door to introducing magnetism into classical electrodynamics.

A good way to visualize an electric dipole is as a small, uniformly-polarized cylinder filled with permanent polarization $P$, lined-up along the cylinder axis; see Fig. 3. The bound charges appear at the top and bottom of the cylinder. It is then easy to see how an externally applied $E$-field pulls and pushes on the top and bottom surfaces of the cylinder. The contributions of the two surfaces to the total force cancel each other out if the $E$-fields acting on the two surfaces happen to be identical. However, if the $E$-field varies from one end of the cylinder to the other, there will be a net force. In the $(P \cdot \nabla)E$ formulation, one only observes this net force; the internal stress that is built-up inside the cylinder is not part of the $(P \cdot \nabla)E$ formula. In contrast, the $-(\nabla \cdot P)E$ formula contains two contributions to the force on the cylinder, one on the top facet, the other on the bottom facet. When these two contributions are added up, their common parts cancel out; what remains is the same as the $(P \cdot \nabla)E$ force on the cylinder. However, the internal stress induced by the $-(\nabla \cdot P)E$ forces is part and parcel of this particular formulation, which is absent from the $(P \cdot \nabla)E$ formulation.

In the vast literature of radiation pressure, just about every paper that analyses radiation forces on material media uses the Einstein-Laub force-density equation. Almost everyone discusses non-magnetic media, so the only part of the Einstein-Laub formula that shows up in these papers is the part that is related to electric polarization $P$. The standard formula for the EM force-density used in radiation pressure literature is $\boldsymbol{F} = (\boldsymbol{P}\cdot\boldsymbol{\nabla})\boldsymbol{E} + (\partial\boldsymbol{P}/\partial t)\times\boldsymbol{B}$, where $\boldsymbol{B} = \mu_0\boldsymbol{H}$—recalling that magnetization is absent. In a 2004 paper,[51] we pointed out that this expression does *not* correspond to the standard Lorentz force-density; throughout that paper we used $\boldsymbol{F}_L = -(\boldsymbol{\nabla}\cdot\boldsymbol{P})\boldsymbol{E} + (\partial\boldsymbol{P}/\partial t)\times\boldsymbol{B}$ and showed some of the "radical" predictions for Lorentz electric dipoles—as compared to Einstein-Laub dipoles which everyone else had been assuming.

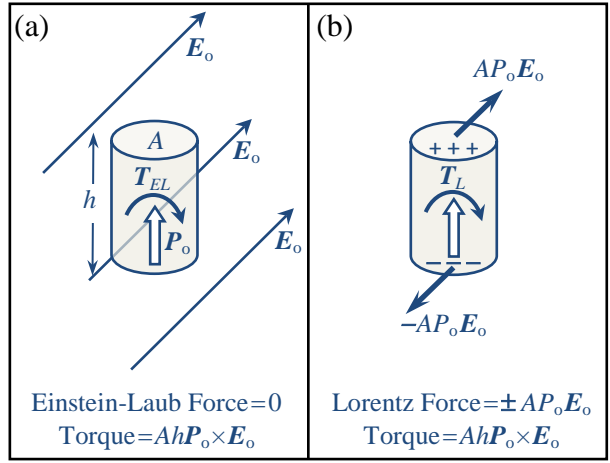

**Fig.3**. A constant and uniform $E$-field $\boldsymbol{E}_o$ acts on a permanently polarized solid cylinder of cross-sectional area $A$, height $h$, and uniform polarization $\boldsymbol{P}_o$. (a) In the Einstein-Laub formulation, the force-density $(\boldsymbol{P}\cdot\boldsymbol{\nabla})\boldsymbol{E}$ is zero everywhere and the torque-density is given by $\boldsymbol{P}_o\times\boldsymbol{E}_o$. (b) In the Lorentz formulation, bound charges with density $-\boldsymbol{\nabla}\cdot\boldsymbol{P}$ appear on the top and bottom facets of the cylinder, giving rise to surface charge densities $\pm P_o$. Thus the force exerted by the external $E$-field on these facets is $\pm AP_o\boldsymbol{E}_o$, even though the net force acting on the cylinder is zero. The torque-density is now given by $\boldsymbol{r}\times\boldsymbol{F}_L(\boldsymbol{r},t)$, which is, once again, equal to $\boldsymbol{P}_o\times\boldsymbol{E}_o$.

Essentially all experimental confirmations of classical radiation pressure involve materials containing electric dipoles and are meant to affirm *not* the putative Lorentz force formula but the Einstein-Laub equation. Interestingly, authors do not seem to be aware that the commonly-used formula is that of Einstein and Laub. We have not been able to trace the origins of this formula's widespread use in the literature; some say it was first used by Landau and Lifshitz.[10] Apparently authors did not go back far enough in time to see it in the 1908 paper of Einstein and Laub.[50]

The Ashkin-Dziedzic experiment involving a focused laser beam on the surface of pure water observed a bulge under the focused spot.[52] This bulge has been shown by Loudon[53] to arise from the Einstein-Laub equation—even though Loudon associates the formula with the name of Lorentz. In fact, the use of the proper Lorentz formalism in conjunction with Maxwell's equations does *not* reproduce the observed bulge.[54] This is one of several reasons why we believe that electric dipoles in material media (water in particular) obey the Einstein-Laub law of force.

**8. A comparison of Chu and Einstein-Laub formulations**. In the Chu formulation,[55] $\boldsymbol{P}(\boldsymbol{r},t)$ is taken to be a continuous, differentiable function, and $-\boldsymbol{\nabla}\cdot\boldsymbol{P}$ acts as a continuous charge distribution, responding to the $E$-field in exactly the same way as does $\rho_{\text{free}}$. In other words, there is no distinction between free-charge and the bound charge of electric dipoles. The dipoles have no individuality; the positive end of one dipole cancels the charge on the negative end of its neighbor, resulting in no force on either dipole. Also, magnetic dipoles are treated similarly to electric dipoles, with the $H$-field acting on magnetic charges and $\varepsilon_0\boldsymbol{E}$ acting on magnetic currents. The force-density and torque-density expressions in the Chu formulation are[12,18]

$$\boldsymbol{F}_{\text{Chu}} = (\rho_{\text{free}} - \boldsymbol{\nabla} \cdot \boldsymbol{P})\boldsymbol{E} + \left(\boldsymbol{J}_{\text{free}} + \frac{\partial \boldsymbol{P}}{\partial t}\right) \times \mu_0 \boldsymbol{H} - (\boldsymbol{\nabla} \cdot \boldsymbol{M})\boldsymbol{H} - \frac{\partial \boldsymbol{M}}{\partial t} \times \varepsilon_0 \boldsymbol{E}, \tag{18a}$$

$$\boldsymbol{T}_{\text{Chu}} = \boldsymbol{r} \times \boldsymbol{F}_{\text{Chu}}. \tag{18b}$$

With the aid of Maxwell's macroscopic equations, the force-density of Eq.(18a) can be shown to be associated with Abraham's EM momentum-density given by Eq.(17b) and the following stress tensor:[12]

$$\overleftrightarrow{\mathcal{T}}_{\text{Chu}}(\boldsymbol{r}, t) = \tfrac{1}{2}(\varepsilon_0 \boldsymbol{E} \cdot \boldsymbol{E} + \mu_0 \boldsymbol{H} \cdot \boldsymbol{H})\overleftrightarrow{\boldsymbol{I}} - \varepsilon_0 \boldsymbol{E}\boldsymbol{E} - \mu_0 \boldsymbol{H}\boldsymbol{H}. \tag{19}$$

Considering that, in free space, the stress tensors of Chu, Einstein-Laub, and Lorentz are identical, we conclude that the *total* EM force on a given object surrounded by vacuum would be the same whether that force is obtained by integrating $\boldsymbol{F}_{\text{Chu}}$ or $\boldsymbol{F}_{EL}$ or $\boldsymbol{F}_L - \partial(\varepsilon_0 \boldsymbol{M} \times \boldsymbol{E})/\partial t$ over the volume of the object. A similar conclusion can be reached with regard to *total* torque on the object.[13] The differences between these formulations are thus confined to the *distributions* of force-density and torque-density within material bodies.[49,56] In the case of materials containing free charge, free current and electric dipoles only, the Chu and Lorentz formulations are identical. In such cases, the differences between Chu and Einstein-Laub are the same as those between the Lorentz and Einstein-Laub formulations. The treatment of magnetic dipoles by Chu parallels his treatment of electric dipoles. Neither Chu nor Einstein-Laub require the introduction of hidden momentum in cases where magnetic dipoles are subjected to an $E$-field. As far as comparison with available experimental data is concerned (e.g., appearance of a bulge on the surface of water under a focused laser beam in the Ashkin-Dziedzic experiment[52]), the Chu formulation shares the shortcomings of the Lorentz formulation.

The behavior of electrons in dielectrics and metals in response to external $E$-fields is usually analyzed using the Lorentz oscillator model and the Drude model.[2,5,9] Bound charges are attached to their atoms and oscillate in the presence of an external $E$-field. Free charges are distributed throughout the lattice, but they also respond to the local $E$-field. However, there is a continuity to the free (i.e., conduction electron) charges that is more in line with the Lorentz/Chu model, whereas the bound electrons behave more in the spirit of the Einstein-Laub model. Of course there are charges which are neither entirely free nor rigidly bound to individual atoms (i.e., itinerant electrons). In this latter case, one should let experiments decide the extent to which such itinerant charges behave similarly to free or bound charges. Experiments such as that of Ashkin and Dziedzic[52] (who focused a laser beam on the water surface and observed a bulge on the surface) show that water molecules respond to the $E$-field in a manner consistent with dipoles which exhibit individuality; such experiments are consistent with the Einstein-Laub model.[53] In the case of radiation forces acting on metals, the conduction electrons pick up most of the force exerted by the light beam; in these cases the electrons behave more or less as free electrons, consistent with the Lorentz/Chu model. Also, in nonlinear optics, the observed electrostriction effects[57-60] in a large number of materials are (at least qualitatively) consistent with the Einstein-Laub theory, showing a healthy disregard for the Lorentz/Chu model.

**9. Why Einstein disavowed the Einstein-Laub formulation**. In response to a June 15, 1918 letter from Walter Dällenbach concerning the EM stress-energy tensor, Einstein wrote: "*It has long been known that the values I had derived with Laub at the time are wrong; Abraham, in particular, was the one who presented this in a thorough paper. The correct strain tensor has incidentally already been pointed out by Minkowski.*"[61] We now know, however, that a major difference between the Lorentz and Einstein-Laub formulations is the presence or absence of hidden entities inside magnetic materials. In other words, it can be shown that the *total* force and *total* torque exerted by EM fields on any object are precisely the same in the two formulations, provided that the contributions of hidden momentum to the Lorentz force and torque on magnetic matter are subtracted.[18] Since the vast majority of the experimental tests of the Lorentz law pertain to total force and/or total torque experienced by rigid bodies, these experiments can be said to equally confirm the validity of the Einstein-Laub formulation.

As for Einstein's reasons, no one knows for sure, but one could speculate. A first guess is that Dällenbach was asking Einstein to explain something in great detail. Einstein did not wish to spend the time and wanted to politely avoid writing a long letter. So he basically said: don't bother with my old paper; study Minkowski's paper instead. This was in 1918, by which time Einstein had probably lost interest in his old electromagnetic papers; his concerns had shifted to general relativity and quantum physics at this point. There is in fact no evidence in the literature that anyone had proven Einstein wrong. (There has been mention of a German paper,[62] which presumably presented such evidence, but we did not find any arguments in this paper against the Einstein-Laub formulation.)

A second guess is that Einstein had realized (perhaps on his own) that his strongly-held belief that the force of a magnetic field on conduction currents must be $\boldsymbol{J} \times \mu_0 \boldsymbol{H}$ (rather than $\boldsymbol{J} \times \boldsymbol{B}$) was wrong. A decade earlier, he was absolutely certain that the force on conduction currents must be $\boldsymbol{J} \times \mu_0 \boldsymbol{H}$. This is the argument with which he and Laub start (rather forcefully) the famous 1908 paper in which the Einstein-Laub force equation is presented.[50] Einstein also discussed this problem of force on conduction currents with Arnold Sommerfeld, who was visiting Einstein in Switzerland around 1908, and wrote to Jacob Laub that Sommerfeld had agreed with his reasoning.

In retrospect, Einstein's mistake was that he was *not* including in his calculations the bound-current-density due to magnetization (i.e., $\boldsymbol{J}_{\text{bound}} = \mu_0^{-1} \boldsymbol{\nabla} \times \boldsymbol{M}$). When one includes this bound-current-density in the force calculations, one finds that both the Lorentz force law and the Einstein-Laub formula give the same answer for total force on any given object. Einstein's initial doubts about the Lorentz force were rooted in his analysis of the force on conduction currents. Presumably, once this argument evaporated, he no longer had a good reason to believe that the Lorentz force was wrong. By 1918, everyone was gravitating toward the Lorentz law, which is somewhat simpler than that of Einstein and Laub; there were no compelling reasons to prefer Einstein's more complex force expression. Shockley's discovery of hidden momentum[15-17] did not happen until after Einstein had passed away. One might speculate that this would have given Einstein a strong incentive to re-examine his old ideas. In any case, Einstein is fascinating even when he thinks he has been wrong.

**Acknowledgement**. Most of the material in this paper originated in extensive discussions with David Griffiths and John Weiner throughout 2013. The author is grateful to both for sharing their deep understanding of classical electrodynamics and for posing many insightful questions.